\documentclass[onecolumn,showpacs,preprintnumbers,amsmath,amssymb,12pt]{revtex4}
\usepackage{graphicx}
\usepackage{dcolumn}
\usepackage{bm}
\usepackage[usenames,dvipsnames,svgnames,table]{xcolor}

\makeatletter
\newenvironment{figurehere}
{\def\@captype{figure}}
{}
\makeatother

\begin{document}

\def\e{\mathop{\rm \mbox{{\Large e}}}\nolimits}
\def\im{\mathop{\rm \od{\iota}}\nolimits}
\newcommand{\ts}[1]{\textstyle #1}
\newcommand{\bn}[1]{\mbox{\boldmath $#1$}}
\newcommand{\bc}{\begin{center}}
\newcommand{\ec}{\end{center}}
\newcommand{\be}{\begin{equation}}
\newcommand{\ee}{\end{equation}}
\newcommand{\bea}{\begin{eqnarray}}
\newcommand{\eea}{\end{eqnarray}}
\newcommand{\ba}{\begin{array}}
\newcommand{\ea}{\end{array}}

\title{A NONEXTENSIVE STATISTICAL MODEL OF MULTIPLE PARTICLE BREAKAGE}
\author{O. Sotolongo-Costa, L. M. Gaggero-Sager, M. E. Mora-Ramos}

\affiliation{Facultad de Ciencias, Universidad Aut\'onoma del Estado de Morelos, Av. Universidad 1001, CP 62209 Cuernavaca, Morelos, Mexico}


\begin{abstract}
A time-dependent statistical description of multiple particle breakage is presented. The approach combines the Tsallis non-extensive entropy with a fractal kinetic equation for the time variation of the number of fragments. The obtained fragment size distribution function is tested by fitting some experimental reports.
\end{abstract}


\maketitle

\section{Introduction}

The reduction of particle size can be achieved under many different conditions. In particular, the multiple breakage of crystals occurs during distinct research and technological processes such as, for instance, milling or recirculation loops. It is, in fact, a very complex phenomenon in which  the quantity known as fragment size distribution function (FSDF) is relevant. In consequence, the problem of finding the time-dependent FSDF (TDFSDF) corresponding to an event of solid multiple breakage, as a function of the main macroscopic measurable variables is of key interest in many areas.

For instance, there are experimental reports on breakage rates \cite{Shook1978}, and on size distribution during particle abrasion \cite{Conti1978,Conti1980}. The usual theoretical approach can be found, in a much complete form, in the work by Hill and Ng \cite{Hill1996}, and more recently in the study published by Yamamoto et al. \cite{Yamamoto2012}.

The aim of the present work is to provide an alternative way to determine the TDFSDF in multiple breakage processes as a function of the relevant macroscopic variables. It is based on the use of a nonextensive statistical description, as it has been previously done in modeling the problems of fragmentation \cite{Sotolongo,SotolongoI}, cluster formation \cite{Calboreanu2003} and particle size distribution \cite{Ramer2005}. The original report on this particular version of the statistics is due to Tsallis \cite{Tsallis1988,Tsallis20091,Tsallis20092}, who postulated a generalized form of the entropy that among other things intends to account for the frequent appearance of power-law phenomena in nature.

On the other hand, the time evolution of the fragment distribution is presented as obeying a fractal-like kinetics \cite{SotolongoIII}. Then, the combination of the nonextensive statistics and the fractal kinetics will allow to derive the expressions for the volume and characteristic length of fragments in breakage events. The paper is organized in such a way that the next section contains a detailed derivation of the distribution functions. Then, the section III is devoted to present and discuss the application of these functions in order to fit with available experimental data and, finally the section IV is devoted to the conclusions.

\section{Statistical Model}

If progressive particle fragmentation takes place within a liquid environment, then quantities such as shear rate and viscosity can be some of the macroscopic variables mentioned above. On the other hand, since the multiple particle breakage involves the effect of inertial impact with container walls or with another particles, it is possible to assume that the fragment mass and the shear rate must appear in the distribution function. The effect of attrition also determines the fragment size, so we also need to take into account the influence of viscosity. Besides, both inertial impact and viscosity depend of fragment concentration. All these variables appear in [1] as the main factors governing the FSDF.

The model considers the following quantities

\begin{itemize}
\item $m\,\rightarrow\,$mass of the fragments.
\item $\dot{\gamma}=\nabla v\,\rightarrow\,$shear rate (gradient of the velocity)
\item $\eta\,\rightarrow\,$liquid viscosity.
\item $n\,\rightarrow\,$number of fragments per unit volume.
\end{itemize}

If the basic dimensions entering this problem are mass ($M$), length ($L$), and time ($T$) (all positive, as noticed); then, according to the Vaschy-Buckingham theorem \cite{Pi1914}, the law relating all these variables can be transformed to one which will include only one dimensionless variable. This provides a method for computing sets of dimensionless parameters for the given variables even if the form of the equation is still unknown.\}
\bigskip

Let us look at this statement core closely.The dimensions involved in the variables above listed are: $m\,\rightarrow\;M$; $\dot{\gamma}\,\rightarrow\;T^{-1}$; $\eta\,\rightarrow\;ML^{-1}T^{-1}$; $n\,\rightarrow\;L^{-3}$.
Accordingly, a single dimensionless quantity defined from them could be

\begin{equation}
\xi =\frac{m\dot{\gamma} n^{1/3}}{\eta};
\end{equation}

Thus, the problem of finding the distribution of volume (mass), or FSDF, can be formulated as the derivation of the distribution function of the dimensionless  variable $\xi$. This can be accomplished using basic Physics principles such as the Second Law of Thermodynamics, which is nothing but a maximization of the system's entropy.

However, the breakage is a phenomenon with long-range correlation among different parts of the system, and the use of the Boltzmann-Gibbs (BG) entropy is not suitable in this case. When long-range correlation is relevant, it turns out that the use of the so-called Tsallis entropy [2] reveals to be convenient. In its continuous version, the form of this entropy (in units of the Boltzmann constant) is:

\begin{equation}
S_q=\frac{1-\int p^{\,q}(\xi)\,d\xi}{q-1}.
\end{equation}

In this expression, $p$ is the probability density function. The quantity $q$ is known as the "degree of non-extensivity" and, in principle, can take any real value. With the use of the L'H\^{o}pital rule, it is possible to verify that, under the normalization condition

\[ \int p(\xi)\,d\xi=1, \]

\noindent the limit when $q\rightarrow 1$ leads to the BG entropy.

The search for a maximum of $S_q$ must include some constraints. One of them is, precisely, the normalization condition

\begin{equation}
\int_0^A p(\xi)\,d\xi=1,
\end{equation}
which is usual in the analysis of the BG case. In the integration, the upper limit $A$ is the maximal value of the dimensionless variable $\xi$, which corresponds to the maximal value of the mass of the fragments, $M$, under stationary conditions for the remaining parameters. In other words, $A=\frac{M\dot{\gamma} n^{1/3}}{\eta}$.

The second constraint is not that usual. In this case, it is customary to impose the finiteness of the so-called $q$-mean value, also named as the first-order $q$-moment:

\begin{equation}
\langle\langle \xi\rangle\rangle_q=\int_0^A \xi\,p^{\,q}(\xi)\,d\xi,
\end{equation}
and, then, the constrain condition should read
\begin{equation}
\langle\langle\xi\rangle\rangle_q=\mu,\;/\vert\mu\vert <\infty.
\end{equation}

The same limits of integration considered in (3)-(4) apply for the integral in the equation (2). Actually, integration limits should include a minimal fragment size that corresponds to the situation when the breakage process cannot yield fragments of smaller dimensions. However, in order to simplify the treatment, our approach sets the lower integration limit as zero.

Under the constrains mentioned, the problem of finding the maximum of the entropy is no other than a Lagrange multipliers one. So, we define the Lagrange functional

\begin{equation}
L[p]=S_q+\alpha\left ( \int_0^A p(\xi)\,d\xi-1 \right )+
\beta\left ( \int_0^A \xi\,p^{\,q}(\xi)\,d\xi-\mu \right ),
\end{equation}
and demand the fulfillment of
\[ \frac{\delta L}{\delta p}=0 \]

The result for the probability density function has the form

\begin{equation}
p(\xi)=\alpha^{\frac{1}{q-1}}\left ( \frac{q}{q-1}\right )^{-\frac{1}{q-1}}[1-\beta (q-1)\xi]^{-\frac{1}{q-1}}.
\end{equation}

Once the Lagrange multipliers $\alpha$ and $\beta$ are properly determined, the final expression for $p(\xi)$ is:

\begin{eqnarray}
p(\xi)&=&\left ( \frac{2-q}{\mu}\right )^{\frac{1}{2-q}}
\left [ 1-\frac{q-1}{2-q} \left (\frac{2-q}{\mu}\right )^{\frac{1}{2-q}}\xi\right ]^{-\frac{1}{q-1}}\!\!\!\!\!;\\
p(\xi)&=&\left ( \frac{2-q}{\mu}\right )^{\frac{1}{2-q}}\left [1-\frac{\xi}{A}\right ]^{-\frac{1}{q-1}}.
\end{eqnarray}

\subsection{The Kinetic Approach}

 During the process of continuous breaking, the concentration of fragments of a given size varies. Therefore, we must consider $\xi= m\dot{\gamma}n^{1/3}/\eta$ as a time-dependent variable. This dependence can be considered as if the fragments were a given species originated during the process. Therefore, the problem will be to determine the kinetics of the fragments.

 In a complex system, a very general kinetic equation for a given species can be posed in the form of a "fractal" differential equation \cite{SotolongoIII}:

 \begin{equation}
 \frac{dN}{dt^{\nu}}=\kappa N,
 \label{kin}
 \end{equation}
where $\kappa$ is the reacting coefficient and $\nu$ is a fractional time index. The solution of the equation (\ref{kin}) is

\begin{equation}
N=N_0 e^{\kappa t^{\nu}},
\label{evol}
\end{equation}

\noindent and defines a kind of "Weibull kinetics", and is a result of our conjecture about the variation of the total number of fragments. If we name $V$ as the total volume of the system, then the density of the fragments of the N species is

\begin{equation}
n=\frac{N}{V}.
\end{equation}

If we set $n_0$ as the initial concentration, $\rho$ the crystal density, and $\tau$ the volume of the crystal fragment, it is possible to write

\begin{equation}
\xi=\frac{\rho\tau\dot{\gamma}n_0}{\eta}^{\!\!\!1/3},
\label{chi}
\end{equation}
and, in correspondence, the time evolution of our dimensionless variable will be -according to (\ref{evol}):

\begin{equation}
\xi(t)=\frac{\rho\tau\dot{\gamma}n_0^{1/3} e^{at^{\nu}}}{\eta}
\label{chi-t}
\end{equation}

Within this context, the time-dependent probability density distribution function for the volume of the fragments can be written as:

\begin{eqnarray}
p(\tau)=\Omega e^{at^{\nu}} [1-B\tau e^{at^{\nu}}]^c;\mbox{\hspace{5.6cm}}\\
{\rm where}\mbox{\hspace{14.7cm}}\nonumber\\
\Omega=\left ( \frac{2-q}{\mu} \right )^{\frac{1}{2-q}} \frac{\rho\dot{\gamma}n_0}{\eta}^{\!\!\!1/3}; \mbox{\hspace{5.8cm}}\\
B=\frac{q-1}{2-q} \left (\frac{2-q}{\mu}\right )^{\frac{1}{2-q}}\frac{\rho\dot{\gamma}n_0}{\eta}^{\!\!\!1/3}; \mbox{\hspace{4.6cm}}\\
c=\frac{1}{1-q}. \mbox{\hspace{8.3cm}}
\end{eqnarray}

\bigskip

It is possible to notice that, written in the form (\ref{chi-t}), the time-dependent probability density function explicitly depends on the macroscopic measurable variables of the system. On the other hand, the integration of (\ref{chi-t}) over the crystal volume, from zero to $\Gamma$, defines the fraction of fragments with a volume size smaller that $\Gamma$, in the system:

\begin{eqnarray}
F(\Gamma)=\int_0^\Gamma p(\tau)d\tau.
\end{eqnarray}

If the volume of the largest fragment, $\Gamma_{max}$, is taken as the unity, one may clearly see that $F(\Gamma_{max})=1$; because any particle in the system will have a volume smaller than the maximal one. This provides a condition for the normalization of the distribution $F(1)=1$, so one obtains

\begin{equation}
F(\tau)=\frac{1-[1-B\tau e^{at^\nu}]^{\frac{2-q}{1-q}}}{ 1-[1-B e^{at^\nu}]^{\frac{2-q}{1-q}}}
\label{FV}
\end{equation}

The distribution (\ref{FV}) gives the fraction of the crystal particles that, at the time $t$, have a volume smaller or equal to $\tau$.

Now, if we look for a distribution in terms of the particle's characteristic length, $l$, instead of volume, we state the cubic dependence of the particle volume with $l$ as $\tau=\sigma l^3/3$. Accordingly, $d\tau=\sigma l^2 dl$. Then,

\begin{equation}
p(l,t)=\sigma \Omega l^2 e^{\kappa t^\nu} [1-\frac{1}{3}\sigma B l^3 e^{\kappa t^\nu}]^{\frac{1}{1-q}}.
\label{dens}
\end{equation}

Again, the length distribution function is defined as the integral of $p(l)$ from zero to a given $l$. A procedure analogous to that previously described leads to the normalized length distribution

\vspace{0.5cm}
\begin{figurehere}
    \centering
 \resizebox{1.0\columnwidth}{!}{\includegraphics[width=0.6\textwidth]{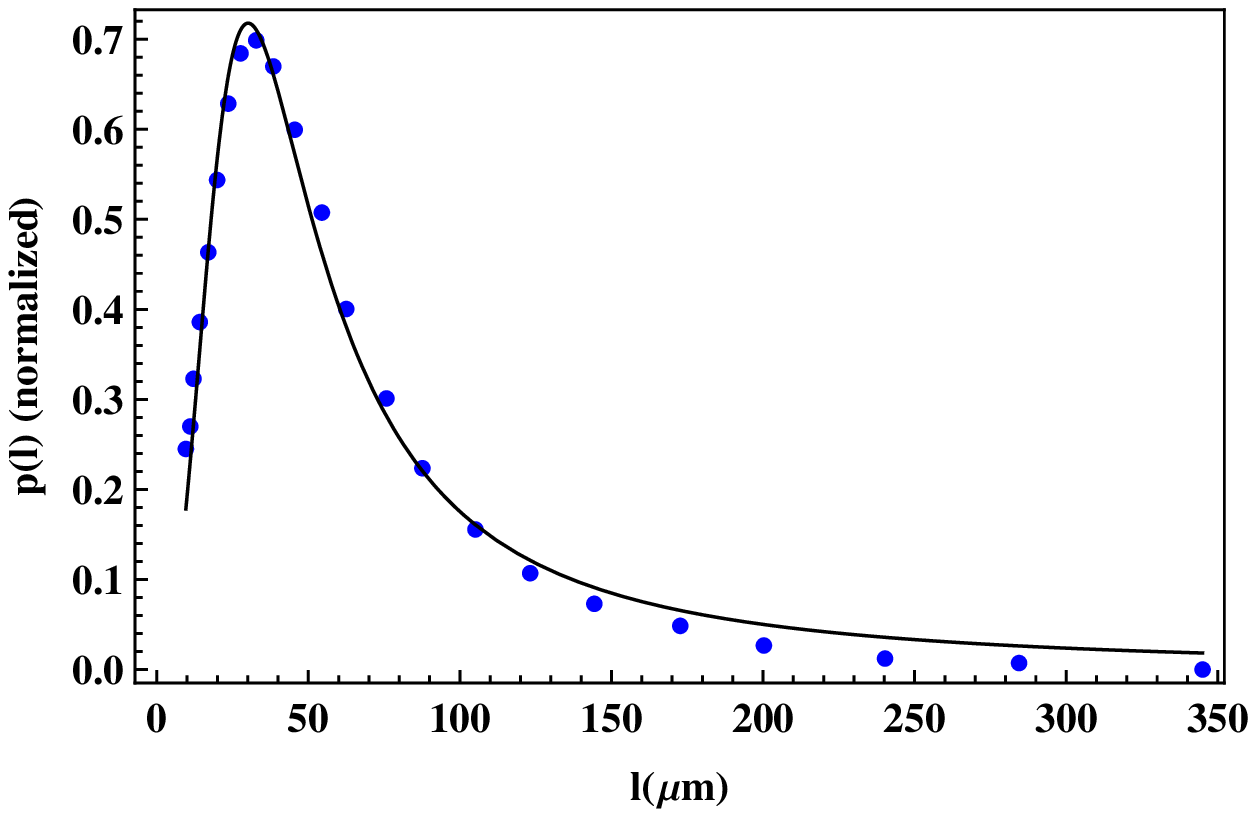}}
    \caption{(color online) Available data on particle size length for phosphate minerals as determined by a commercial analyzer \cite{Outotec} (dots), and best static fit obtained with the use of Eqn. (28) (solid line).}
    \label{F55}
\end{figurehere}
\vspace{0.5cm}

\begin{equation}
f(l,t)=\frac{1-[1-\sigma B l^3 e^{\kappa t^\nu}]^{\frac{2-q}{1-q}}}{1-[1-\left ( \frac{1}{\sigma^2}\right ) B e^{\kappa t^\nu}]^{\frac{2-q}{1-q}}}.
\end{equation}

\section{Results and discussion}

In order to test our model we fit the results of the particle diameter distribution of phosphate minerals reported as an outcome of a commercial particle size analyzer (see Ref. \cite{Outotec}). Since the experimental information is not time-depending, we chose a stationary fitting procedure that uses the expression derived for the -normalized- distribution density in such a way that the proposed expression will contain three adjusting parameters, $\{r,\,s,\,q\}$. The choice of $r$ and $s$ is due the presence of such unknown quantities as $\mu$, $\sigma$, and the time-related exponents in the expression. Then;

\begin{equation}
p(l)=r\,l^2\,(2-q)^{\frac{1}{2-q}}\,\left [1-s\,l^3\,(2-q)^{\frac{1}{2-q}}\,\left (\frac{q-1}{2-q} \right )\right ]^{\frac{1}{1-q}}.
\label{fit}
\end{equation}

The normalized data and the resulting curve appear in the Fig. 1. We used the Wolfram Mathematica "NonLinearModelFit" package with the Finite Difference Gradient Method and the sole restrictions $r>0$, $s<0$ and $1<q<2$. The best fit parameters obtained are: $r=1.94$, $s=-0.01$, and $q=1.78$, with the goodness of the fit determined by the parameter $R^2$ with a value of $0.995$. It is worth to remark that the particular value for the Tsallis q-parameter in this fragment size distribution is below the $q=1.97$ reported by Calboreanu et al. \cite{Ramer2005}.

Conti and Nienow \cite{Conti1980} reported on the abrasion experiments in a solution with a solid phase of nickel-ammonium sulphate hexa-hydrate crystals. The procedure included stoping the agitation at certain intervals. Then, the abrasion fragments were separated from the liquid to determine the total mass abraded and to measure size distribution. The counted particles were grouped into a number of size ranges in order to give a clearer representation of the changes of size distribution with time. We have, for instance: $0-36\,\mu$m (range 1); $36-73\,\mu$m (range 2); $73-109\,\mu$m (range 3). Assuming a constant shape factor, the total particle mass in each of the ranges was calculated (see both Table 1 and Figure 1 in Ref. \cite{Conti1980}).

We use the kinetic model proposed in this work for fitting the mentioned results in Ref. \cite{Conti1980} of the time-dependent distribution of abraded particle mass (size). The expression used to fit the different data sections is derived from (\ref{fit}) considering that the exact values of the particle size in each data point are not known and will be considered as parts of the adjustment. This leads to consider the exponents in the time-evolution law (\ref{evol}) as new fitting parameters, $\{u, v\}$, through  the substitutions $r\,l^2\rightarrow r\exp{(u\,t^v)}$ and $s\,l^3\rightarrow s\exp{(u\,t^v)}$; considering $u$ and $v$ as another two fitting parameters;

\begin{equation}
p(t)=r\,(2-q)^{\frac{1}{2-q}}\,e^{u\,t^v}\left [1-s\,(2-q)^{\frac{1}{2-q}}\,\left (\frac{q-1}{2-q} \right )\,e^{u\,t^v}\right ]^{\frac{1}{1-q}}.
\label{fitT}
\end{equation}

The figure 2 contains the results of the fitting of the total mass of the fragments (curve and dots in black) as well as the fitting of the data corresponding to the above mentioned three first particle size (mass) ranges appearing in Table 1 of Conti and Nienow report (in blue, red, and purple color, respectively). The fitting was carried out with the same package

\vspace{0.5cm}
\begin{figurehere}
    \centering
 \resizebox{1.0\columnwidth}{!}{\includegraphics[width=0.6\textwidth]{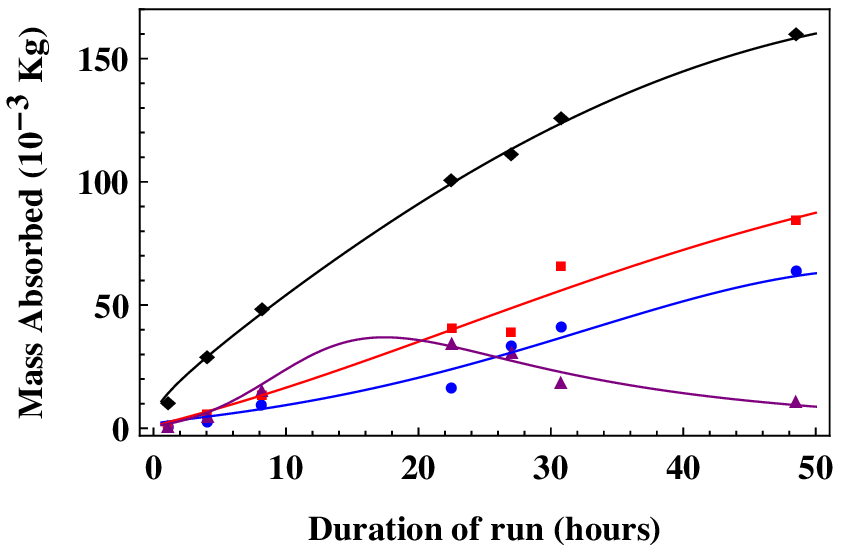}}
    \caption{(color online) Available data on particle mass (size range) for nickel-ammonium sulphate hexa-hydrate crystals \cite{Conti1980} as functions of abrasion time (dots); and best kinetic fit obtained with the use of Eqn. \ref{fit}, modified along the comments in the text (solid lines). Black curve (diamonds) corresponds to the measured total mass of the fragments. Mass distribution of particles within range 1 (blue); range 2 (red); range 3 (purple).}
    \label{F55}
\end{figurehere}
\vspace{0.5cm}

\noindent above referred. The conditions for the fitting procedure were set as: $r>0$, $s<0$ (in order to keep the distribution as a real quantity), $t>0$, $u>0$, and $1<q<2$. In all cases the characteristic fitting index $R^2$ ranked above $0.985$.

The values obtained for the distinct parameters are the following. Total mass (black, diamonds): $r=3.63$, $s=-0.006$, $q=1.27$, $t=1.79$, $u=0.26$. First size range (blue, circles): $r=1.97$, $s=-0.01$, $q=1.13$, $t=0.48$, $u=0.57$. Second range (red, squares): $r=2.12$, $s=-0.004$, $q=1.63$, $t=2.80$, $u=0.24$. Third range (purple, triangles): $r=107.52$, $s=-0.47$, $q=1.74$, $t=1.06$, $u=0.60$.

It must be stressed that the values of $\nu$-exponent here obtained constitute a validation of our hypothesis of fractal kinetics to describe the time evolution of the FSDF. On the other hand, it is possible to notice that a good fitting can be achieved by demanding that the non-extensivity parameter be a non-integer with values between $1$ and $2$, as previously found. One readily notices that, the highest values of $q$ obtained are around $1.7$, which is quite below the value of $2$, and the value of $1.97$ reported in Ref. \cite{Ramer2005}.

\section{Conclusions}

The use of a non-extensive statistical description is a correct choice for deriving the fragment size distribution function arising from processes of multiple breakage of crystals in stirred vessels. This can be performed both under static and time-dependent conditions, with the introduction of a suitable fractal kinetics, and fractional power time evolution.

The proposed distribution functions were tested by fitting available experimental data. We have shown that the statistics of multiple breakage phenomenon can be modeled via a Tsallis entropy. On the other hand, we have obtained values of the magnitude of the fractal time exponent within the range between $0.2$ and $0.6$.

\end{document}